\documentclass[usenatbib]{mn2e}
\usepackage{multirow}

\newcommand{\apjl}{ApJL}
\newcommand{\aap}{A\&A}

\newcommand{\apjs}{ApJS}

\usepackage{epsfig,amsmath,enumerate,amssymb,aas_macros,bm,booktabs,multirow,xcolor}
\usepackage{graphics}

\newcommand{\etgs}{{ETGs}}
\newcommand{\etg}{{ETG}}

\newcommand{\solarmass}{ \rm M_\odot}

\newcommand{\omegam}{\Omega_{\rm m}}
\newcommand{\omegal}{\Omega_{\rm \Lambda}}
\newcommand{\omegab}{\Omega_{\rm b}}

\newcommand{\reff}{R_{\rm e}}
\newcommand{\mstar}{M_{\rm \star}}

\def\gsim { \lower .75ex \hbox{$\sim$} \llap{\raise .27ex \hbox{$>$}}}
\def\lsim { \lower .75ex \hbox{$\sim$} \llap{\raise .27ex \hbox{$<$}}}
\begin{document}

\title{The Size Evolution of Elliptical Galaxies}
\author[Xie et al.] {
\parbox{0.9\textwidth}{Lizhi Xie$^{1}$\thanks{Email:lzxie@bao.ac.cn}, 
        Qi Guo$^{1,2}$,
	Andrew P. Cooper$^{1,2}$,
	Carlos S. Frenk$^{2}$,
	Ran Li$^{1}$,
	Liang Gao$^{1,2}$}
\vspace*{4pt} \\
$^1$Key Laboratory for Computational Astrophysics,
The Partner Group of Max Planck Institute for Astrophysics,\\
National Astronomical Observatories, Chinese Academy of Sciences,
 Beijing, 100012, China\\
$^2$Institute of Computational Cosmology, Department of Physics,
University of Durham, Science Laboratories,\\
 South Road, Durham DH1 3LE \\
}

\maketitle
\label{firstpage}

\begin{abstract}

Recent work has suggested that the amplitude of the size mass relation of
massive early type galaxies  evolves with redshift.  Here we use a
semi-analytical galaxy formation model to study the size evolution of massive
early type galaxies. We find this model is able to reproduce the amplitude of
present day amplitude and slope of the relation between size and stellar mass
for these galaxies, as well as its evolution.  The amplitude of this relation
reflects the typical compactness of dark halos at the time when most of the
stars are formed. This link between size and star formation epoch is propagated
in galaxy mergers.  Mergers of high or moderate mass ratio (less than 1:3)
become increasingly important with increasing present day stellar mass for
galaxies more massive than $10^{11.4}M_{\odot}$. At lower masses, low mass
ratio mergers play a more important role.  In situ star formation contribute
more to the size growth than it does to stellar mass growth.  We also find
that, for ETGs identified at $z=2$, minor mergers dominate subsequent growth
both for stellar mass and in size, consistent with earlier theoretical results.

\end{abstract}

\section{Introduction}

The most massive galaxies are typically early type galaxies (hereafter ETGs).
Understanding the abundance and properties of \etgs is very important for
galaxy formation theory and is also relevant to the determination of
cosmological parameters \citep[e.g.][]{delucia2007, conselice2014}.
 The evolution of ETGs is thought to be driven by mergers, and hence to reflect the
 hierarchical nature of structure formation in the $\Lambda$CDM model
 \citep{frenk1985,white1991,kauffmann93,lacey93,parry09}. 

Size is one of the most important observables in efforts to understand the evolution of ETGs. The size of a galaxy is typically defined as the projected radius, $\reff$, containing half of its stellar mass, $\mstar$. The scaling relation between $\reff$ and $\mstar$ for the ETG population has been studied by a number of recent galaxy surveys [e.g.\citep{shen03,bernardi2010,buitrago08,vander14,cassata13,cooper2012}].  It has been known for many years that ETGs (defined according to various combinations of mass, star formation rate, colour and surface brightness profile shape) are much more compact at high redshift, compared to their counterparts in the local Universe \citep{Daddi05, mcintosh05, trujillo06, diSergo05}. An often-quoted result is that the effective radius of a `typical massive ETG' increases by up to a factor of $\sim4$ from $z = 2.5$ to $z = 0$ \citep{trujillo06, cassata11, cenarro09, buitrago08, vandokkum08, damjanov11, vander14}. Over the same redshift range, the corresponding `typical mass' increases by only a factor of 2 \citep{vandokkum10, baldry12}. These results most often refer to the average size and mass of all ETGs above a fixed mass, applying the same rest-frame selection criteria at all redshifts. Recent work has provided more detailed insights: for example, there is evidence that the size increase may have been be larger for galaxies of larger present-day stellar mass \protect\citep{ryan12}, and \cite{saracco2011,shankar13,napolitano2010} found that the size of ETGs selected may depend on their stellar age. This age dependence is not apparent in the local Universe \citep{trujillo11}. The central stellar mass density of massive galaxies at high redshift is similar to that of comparable galaxies in the local Universe \citep{bezanson2009,Hopkins2009,vandokkum10,huang13}, and the majority of the evolution of the stellar mass density profiles of these galaxies seems to occur in their low surface brightness outer regions \citep{saracco2012}. Similarly, the central velocity dispersions of ETGs shows only a weak decline with decreasing redshift at $z \lesssim 2$ \citep{cenarro09}.

Explanations for these observations have been sought in the context of the $\Lambda$CDM model. This is far from straightforward -- unlike the growth of dark matter structure, galaxy evolution in $\Lambda$CDM has a number of redshift-dependant characteristic scales introduced by baryonic physics \citep{guo08}.  Moreover, when galaxy populations are defined by redshift-independent selection functions (as in the case of ETGs), it becomes necessary to account for the apparent `evolution' due to galaxies entering and leaving these selections, in addition to the evolution of galaxies that remain in the sample from high to low redshift. A fully self-consistent theory of how the ETG population evolves therefore requires a complete forward model of the entire galaxy population.

Here we study the size-mass relation of the ETG population at different epochs, from $z\sim2$ to $z=0$, in the semi-analytic galaxy formation model developed by \cite{guo11, guo13}. This model reproduces many properties of galaxies observed in the local Universe and at high redshift, including the size -- mass relation for both early and late type galaxies at the present day \citep{guo11}. We compare our model to observed galaxy size data over the same redshift range. In the context of the evolving amplitude of the size -- mass relation, we examine the origin and relative importance of sample evolution and intrinsic evolution, concentrating on the mechanisms naturally provided by standard galaxy formation theory, namely star formation and dissipationless merging.

Mergers of high mass ratio, in particular, are thought to be capable of
increasing galaxy size while providing relatively little corresponding increase
in mass and having little or no effect on central density or velocity
dispersion
\citep{cole2000,hilz2013,Hopkins2009,trujillo11,newman2012,bezanson2009}.
Evolution dominated by these `minor' mergers  therefore provides a plausible
explanation of the observational results at $z\lesssim2$ mentioned above
[$\delta \reff \propto (\delta \mstar)^2$,  e.g. \citealt{naab09}].
Observational arguments supporting this hypothesis have also been made based on
the greater frequency of higher mass ratio mergers
\citep[e.g.][]{trujillo13,mclure13}.  Recent hydrodynamical simulations
\citep[e.g.][]{naab09,oser12} and N-body experiments \citep{laporte13} have
demonstrated that this explanation is indeed plausible in a cosmological
context. In our model, the various evolutionary processes relevant to ETGs are
included consistently with one another and with the galaxy population as a
whole, allowing us to comment further on the relative importance of minor
mergers. Note we do not address the nature of so-called ultra-compact
galaxies, which represent a small fraction of the $z\sim 2$ ETG population and
are thought to form through intense, highly dissipative starbursts
\citep[e.g.][]{Dekel2009, Hopkins2009}.

This paper is structured as follows. In section~\ref{sec:simu} we briefly describe the N-body simulation and semi-analytic model used for this work. In section:~\ref{sec:size} we compare size--mass relations at different redshifts in the model to observed relations. In Section~\ref{sec:grow} we study the different mechanisms driving the evolution of the size -- stellar mass relation in the model. We summarize our results in Sec.~\ref{sec:conclusion}.    

\section{Simulation and semi-analytical models}
\label{sec:simu}

The galaxy formation model in this work is based on dark matter halo merger trees extracted from the cosmological $N$-body {\it Millennium Simulation}. Descritpions of the {\it Millennium Simulation} and our galaxy formation model can be found in \cite{springel05} and \cite{guo11, guo13}, respectively. Here we summarize the most important characteristics of the simulation and the equations in the model relevant to the sizes of ETGs.  

\subsection{The simulation}

The {\it Millennium Simulation} \citep{springel05} is a cosmological N-body simulation, which follows $2160^{3}$ particles from redshift $z=127$ to the present day in a box of length 500 Mpc/$h$ on each side (where the Hubble parameter $h=0.73$). This volume is large enough to investigate the statistical distributions of the properties of massive ETGs. Each dark matter particle has a mass of $8.6\times 10^8
M_{\odot}/h$, allowing us to follow galaxies down to masses comparable to that of the Small Magellanic Cloud. The simulation adopted cosmological parameters consistent with the first year WMAP results: $\omegam =0.25, \omegab=0.045, \omegal =0.75, \sigma_{8}=0.9, n=1$.

Particle data were stored at 64 logarithmically spaced output times. At each snapshot, the Friends-of-Friends groupfinding algorithm was used to link particles separated by less than 0.2 of the average interparticle separation \citep{Davis1985}. The SUBFIND algorithm \citep{springel01} was then applied to decompose these groups into self-bound substructures (hereafter subhaloes). Merger trees were constructed by linking subhalos at different output times into chains of progenitors and descendants using the algorithm described in \cite{springel05,boylankolchin09}. The galaxy model then processes these merger trees.

We call the most massive subhalo in a FOF group the `main halo'. A galaxy assigned to the potential minimum of a main halo is referred to as a central galaxy, while galaxies assigned to satellite subhalos are referred to as satellite galaxies. Satellite galaxies include so-called `orphans' whose subhalos cannot be resolved anymore by the $N$-body simulation; the orbits of these galaxies are tracked semi-analytically, such that the ability to follow satellites until they merge is not limited by the resoultion of the $N$-body simulation \citep{springel01}. For the main halo of each FOF group, we define a total mass, $M_{\rm 200}$, enclosed by a radius, $R_{\rm 200}$, within which the mean density is 200 times the critical density for closure at the corresponding redshift.

\subsection{Semi-analytical Model}

In the standard $\Lambda$CDM model, galaxies grow in dark matter dominated potentials as the result of in situ star formation in condensed gas and the accretion of less massive satellite galaxies \citep{whiterees}. In contrast to the self-similar growth of dark matter halos, the rate of change of stellar mass through both of these channels varies according the existing stellar mass and, at a fixed mass, with redshift \citep{guo08}. In this model, star formation always dominates the growth of stellar mass for low mass galaxies (present-day $M_{\star} \lesssim 10^{10} \, M_{\odot}$). More massive galaxies grow mainly through star formation at $z \gtrsim 2$ and by accretion and merging thereafter. Detailed descriptions of stellar mass growth in similar galaxy formation models can be found in \cite{guo11, guo13, delucia2007, croton06, springel01}. This model also tracks changes in the size of galaxies as their mass evolves. Galaxies are separated into three components -- gas disks, stellar disks and stellar bulges. The size and mass of each of these components is followed separately, according to the following prescriptions. 

\subsubsection{Disk sizes}

When gas condenses into the centre of a potential well, we assume that it has the same specific angular momentum ($\textbf{j}_{gas,cooling}$) as its  host halo. The total angular momentum of the gas disk is thus \begin{equation}
  \textbf{J}_{gas,new} = \textbf{J}_{gas,old} + \textbf{J}_{gas,cooling}- \textbf{J}_{gas, SF} +\textbf{J}_{gas,merger},
\end{equation} where $\textbf{J}_{gas,new}$ and $\textbf{J}_{gas,old}$ are the new and original total angular momentum of the gas disk, respectively. $\textbf{J}_{gas,cooling}$ is the total angular momentum of recently cooled gas ($\textbf{J}_{gas,cooling} = m_{gas, cooling}\textbf{j}_{gas, cooling}$, where $m_{gas, cooling}$ is the amount of gas cooled in a given time interval) and $\textbf{J}_{gas,merger}$ is the total angular momentum carried by the gas component of merging satellites. $\textbf{J}_{gas, SF}$ is the angular momentum lost to stellar disk through star formation.

The stellar disk gains angular momentum through star formation and loses it through disk instabilities, which transfer stars from the disk to the bulge component as required to marginally stabilize the stellar disk against gravitational instability. The balance equation for the stellar disk angular momentum is therefore \begin{equation}
  \textbf{J}_{\star,new} = \textbf{J}_{\star,old} + \textbf{J}_{\star, SF}  - \textbf{J}_{\star,instability} \end{equation} where \begin{equation}
  \textbf{J}_{\star, SF} =  \textbf{J}_{gas, SF} = M_{\star, SF} * \textbf{j}_{gas}.
\end{equation} Here $M_{\star, SF}$ is the amount of new formed star and $\textbf{j}_{gas}$ is the specific angular momentum of the gas disk. 

We assume both the gas disk and the stellar disk have exponential surface density profiles and the circular velocity curve is flat, hence the exponential scale length of the gas or stellar disk is given by \citet{croton06}
 \begin{equation}
 \label{eqn:stardisksize}
 R_{gas,\star} = \frac{\textbf{J}_{gas,\star}/M_{gas,\star}}{2V_{cir}},
\end{equation} where $M_{gas,\star}$ denotes the total mass in the gas or stellar disk and $\textbf{J}_{gas,\star}$ the corresponding total angular momentum. In practice we use $V_{max}$, the maximum circular velocity of the dark halo, as a proxy for $V_{cir}$.
Note this directly connect the sizes of disks to the characteristic scale of their dark matter halo.

\subsubsection{Spheroid sizes}

When halos become subhalos of more massive systems, their galaxies become satellites. Satellites with resolved subhaloes survive until either (i) they are deemed to be tidally disrupted or (ii) their corresponding subhalo is lost from the $N$-body simulation \textit{and} the time for their inspiral to the centre of their host potential is less than the lookback time at which their subhalo was lost. Further caveats to these prescriptions are described in \cite{guo11}. Stars from tidally disrupted objects are placed into a 'stellar halo' reservoir. For the purposes of this paper, these stars are considered to be unobservable, and they are are never transferred back to any central galaxy. Hence only stars from satellites merging to the centre of their host can influence the size of the host's central galaxy.

Binary mergers between galaxies (and dark matter haloes) are often divided into `major' and `minor' categories according to mass ratio of the two progenitors. Following convention in the literature, \cite{guo11} used a baryon mass ratio
threshold of $3:1$ to divide `major' and `minor' mergers. In this model of major mergers, violent relaxation leads to the complete destruction of centrifugally support disks, such that the remnant is purely dispersion-suppoted \citep[e.g.][]{naab07}. In contrast, the disk of the more massive (primary) progenitor is allowed to survive in minor mergers, with the stars from the less massive (secondary) progenitor being scattered into the stellar spheroid (bulge) of the remanant. 

Mergers also trigger rapid gas disipation, represented by a `starburst' mode of star formation. In the case of major mergers, all gas from both progenitors is used to fuel a starburst that adds stars to the spheroidal component of the remnant. In the minor merger case, gas from the secondary progenitor is added to the gas disk of the remnant, and stars formed in the starburst are also added to the surviving stellar disk. These bursts can convert a large fraction of the available cold gas into stars. A recipe from \cite{somerville01} is used here to model the fraction of gas converted into stars during a starburst: \begin{equation}
\epsilon_{bust} = 0.56 \times (\frac{M_{\rm sat}}{M_{\rm cen}})^{0.7}
\end{equation} where $M_{\rm sat}$ and $M_{\rm cen}$ are the total baryonic mass of the satellite and central galaxies, respectively.

Assuming energy conservation and virial equilibrium, the growth in the size of the spheroid component in a merger between two galaxies can be approximated by \citet{cole2000}:
\begin{equation}
 C\frac{GM^{2}_{\rm new,b}}{R_{\rm new,b}} =
C\frac{GM^{2}_{
\rm sat}}{R_{\rm sat}}+C\frac{GM^{2}_{\rm cen}}{R_{\rm cen}}+\alpha\frac{GM_{\rm sat}M_{\rm cen}}{R_{\rm sat}+R_{\rm cen}}.
\label{eqn:bulgesize}
\end{equation} Here $C$ is the so-called structure parameter, relating the binding energy of a galaxy to its mass and radius, and the factor $\alpha$ parmeterizes the effective interaction energy of the two galaxies. 

In major mergers, both existing stars and stars formed in the associated starburst are counted when calculating the spheroid size of the remnant, i.e. $M_{\mathrm{cen/sat}} = M_{\mathrm{cen/sat},\star} + \epsilon_{\rm burst}\times M_{\mathrm{cen/sat,gas}}$ where $M_{\mathrm{cen/sat,\star}}$ and $M_{\mathrm{cen/sat,gas}}$ are the stellar mass and cold gas mass of the central and satellite galaxies, respectively. In minor mergers, all stars in the satellite galaxy are added to the bulge of the central galaxy. The corresponding $M_{\rm cen}$ and $M_{\rm sat}$ are then assumed to be $M_{\rm cen, b}$ and $M_{sat, \star}$, respectively, where $M_{\rm cen, b}$ denotes the stellar mass in the spheroids of the central galaxies. $R_{\rm sat}$ and $R_{\rm cen}$ are the corresponding half mass radii.

Dynamical instabilities in the disk are another important channel by which stars are transfered to the spheroidal component. \citet{guo11} assumed a simple criterion to estimate the onset of instability, 
\begin{equation} V_{\rm max} < \sqrt{\frac{GM_{\rm disk,\star}}{3R_{\rm disk,\star}}},
\label{eq:instable}
\end{equation}
where $M_{disk,\star}$ is the mass of the stellar disk and $R_{disk,\star}$ its exponential scale length.
When Eq.~\ref{eq:instable} is satisfied, a stellar mass of $\delta M_{\star}$ is transfered from the disk to the spheorid such that the disk is made marginally stable. The corresponding growth in spheroid size is also modeled using Eq.~\ref{eqn:bulgesize}, defining $M_{\rm cen}$ and $R_{\rm cen}$ to be the stellar mass and the half mass radius of the existing bulge, if any, and $M_{\rm sat} = \delta M_{\star}$ and $R_{\rm sat}$ to be the radius containing a mass $\delta M_{\star}$ in the unstable disk. The pre-factor $\alpha$ is set to $2$ in this case, higher than in the merger case, since the `old' and `new' spheroid stars at least partly overlap at the onset of the instability, implying a higher interaction energy.

Although the \citet{guo11} model accounts for the rapid conversion of gas disipated in mergers to stars, the energy balance represented by Eqn.~\ref{eqn:bulgesize} does not take this dissipation into account. Doing so would reduce the size of the remnant further in cases where gas makes up a substantial fraction of the mass of either progenitor \cite[e.g.][]{covington2008}. \citet{shankar13} have shown that the sizes of low mass ETGs in this model would be in better agreement with observations if this effect was included. Such changes are beyond the scope of the present paper. 

\subsection{Projected half mass radius}
\label{subsec:phr}

All the radii discussed above are defined in three dimensions. To compare
these to observational data directly, we need to convert them to radii in projection. We assume that the spheroidal components of our galaxies follow a Jaffe profile \citep{jaffe83}: \begin{equation}
j_{\rm b}=\frac{M_{\rm b} r_{\rm{b}}}{4\pi r^2(r+r_{\rm b})^{2}}.
\label{eqn:p_bulgesize3d} 
\end{equation}
Here $M_{\rm b}$ is the stellar mass of the bulge, and $r_{\rm b}$ is the corresponding 3-D half-mass radius. The projected density profile is given by 
\begin{equation}
I_{\rm b}(R)=\int _R^{\infty} j(r)\times \frac{r}{(r^2-R^2)^{1/2}} dr.
\label{eqn:p_bulgesize2d} 
\end{equation} We assume the Jaffe model to simply this calculation, noting that it may not be a good description of all galactic spheroids.

We assume an exponential profile for the disk component, 
\begin{equation}
I_{\rm d}=\frac{M_{\rm d}}{2\pi r_{\rm d}^2} e^{-r/r_{\rm d}},
\label{eqn:p_disksize} 
\end{equation}
where $M_{\rm d}$ is the stellar mass of the disk and $r_{\rm d}$ the exponential scale length.  The total  surface mass density (surface brigtness) is the sum of these two components\begin{equation}
I(r) = I_{\rm b}(r)+I_{\rm d}(r).
\end{equation}

\subsection{Fiducial ETG definition}
\label{subsec:fiducialetg}

 A wide variety of definitions of the ETG population are found in the literature. Some specifiy the fraction of light or mass in the spheroidal component, some impose upper limints on specific star formation rate, and some select by color or spectral shape. To perform a meaningful comparison between model predictions and observational data in the following section, we will adapt our selection criteria to match roughly those of each dataset we compare to. For simplicity, however, in all other sections this paper we use only one fiducial classification of ETGs in the model, according to their total stellar mass, bulge-to-total stellar mass ratio and specific star formation rate: \begin{equation}
M_{\star} > \, 1\times 10^{11}\solarmass, \,\,
M_{\rm b}/M_{\star} > \, 0.9,\,
sSFR < \,\, 10^{-11} \mathrm{yr^{-1}} 
\label{eqn:fiducialetg}
\end{equation}.

\section{Size evolution}
\label{sec:size}

\subsection {Model vs. observation}
\label{subsec:sizeevo}

In this section, we compare the evolution of the size of `typical' ETGs and the entire ETG size--mass relation between model predictions and  observations.

\begin{figure}
\includegraphics[width=0.45\textwidth]{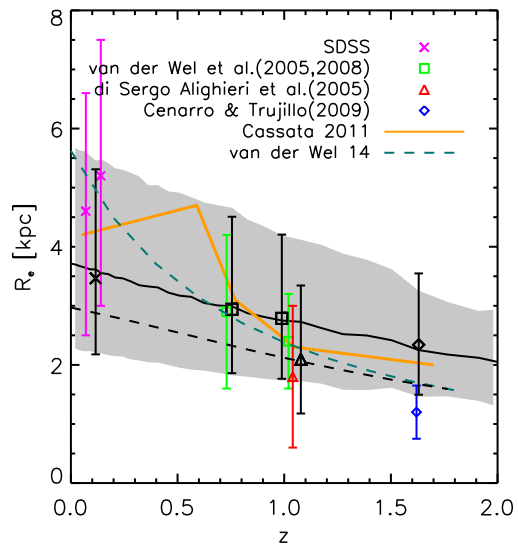}

\caption{Projected half mass radius as a function of redshift. 
Black curves and symbols are the model predictions, while colored curves and
symbols with error bars are the measurements in the literature. Different
observations use somehow different selection criteria for ETGs. Detailed
selection criteria can be found in Table~\ref{tab:select}. Solid, dashed curves and symbols denote the
median values of the corresponding size distribution. Errors along the
y-axis and the shaded region represent 16 per cent to 84 per cent ranges.
} 
\label{fig:radius}
\end{figure}

Fig.~\ref{fig:radius} shows how the median size of the ETG population in the model, defined according to our fiducial criterion (eqn.~\ref{eqn:fiducialetg}), varies with redshift (solid black line). The median size of galaxies selected in this way increases by a factor of $\sim1.8$ between redshift $z\sim2$ and $z=0$. 

We compare this prediction with a number of recent observational estimates obtained at different redshifts. For each coloured point or line in Fig.~\ref{fig:radius}, representing an observational result, there is a corresponding black point or line of the same style representing an equivalent selection from the model. The datasets and selection criteria are summarised in Table~\ref{tab:select}. To reflect uncertainties in the observational determination of stellar mass \citep{longhetti2009,mitchell2013}, we have convolved the stellar mass of each model galaxy with a Gaussian of dispersion 0.25 dex in $\log_{10}\,M_{\star}$. Where necesary, we have recalibrated stellar masses from observations to the assumption of a universal Chabrier IMF \citep{Chabrier03} used by \cite{guo11}.

\begin{table*}\tiny
\caption{A summary of the observational data we compare to the model. From left to right, columns are as follows: [1] observational data source (as 
Fig.~\ref{fig:radius}); [2] the survey or catalog from which the data orginates; [3,4,5,6] respectively mass, morphology, specific star formation rate and redshift criteria that define each sample, in the observational data and the corresponding model selection; [7] the 
symbol/line style and colour denoting the data and its model comparison in Fig.~\ref{fig:radius}. Note that morphological criteria are most difficult ot match with the model. When selecting 
ETGs' morphology, \protect\cite{cassata11,vander05,vander08} selected visually spheroidal galaxies 
; \protect\cite{cenarro09} selected galaxies with Sersic index $n > 2.5$; and
\protect\cite{diSergo05,vander14} did not set limitations on morphology or 
sSFR, but instead  selected galaxies by an early type spectral classification and color, respectively.
}
\centering
\begin{tabular}{lcccccc}
	\toprule
	 former works & surveys & $M_{\star}$ [$ 10^{11}\solarmass$] & morphology & sSFR [$10^{-11}/yr$] & z & symbol/line   \\
	\midrule
	\cite{cassata11} & GOODS WFC3 & $M_{\star} > 1$ & visually spheroidal & $sSFR<1$ & 0<z<2.5 & yellow line \\
	 model & & $M_{star} > 1$ &  $M_{b}/M_{\star} > 0.9$ & $sSFR<1$ & 0<z<2.0 & black line\\
	\midrule
	\cite{cenarro09} & SDSS DR6 & $0.5 < M_{\star} < 2 $ & S0, E & & $0<z<0.1$ & Purple crosses \\
	 model & & $0.5 < M_{\star} < 2$ & $M_{b}/M_{\star} > 0.9$ & & $z \sim 0.12$ & Black cross \\
	\midrule
	\cite{cenarro09} & GMASS & $0.5 < M_{\star} < 2 $ & S0, E & & $1.4<z<2.0$ & blue diamond \\
	 model &  &$0.5 < M_{\star} < 2$ & $M_{b}/M_{\star} > 0.9$ & & $z \sim 1.63$ & Black diamond \\
	\midrule
	\cite{vander05} & CDFS, RDCS & $0.5 < M_{\star} < 2 $ & S0, E & & 0.9<z<1.2 & green squares \\	
	 model &  & $0.5 < M_{\star} < 2$ & $M_{b}/M_{\star} > 0.9$ & & $z \sim 0.99$ & Black square \\
	\cite{vander08} & CDFS, RDCS & $0.5 < M_{\star} <2 $ & S0, E & & 0.6<z<0.8 & green squares \\	
	 model &  & $0.5 < M_{\star} < 2$ & $M_{b}/M_{\star} > 0.9$ & & $z \sim 0.76$ & Black square \\
	\midrule
	\cite{diSergo05} & K20 & $0.5 < M_{\star} < 2 $ & & early-type spectrum & 0.88<z<1.3 & red triangle \\
	 model &  & $0.5 < M_{\star} < 2$ & & $sSFR< 1$ & $z \sim 1.1 $ & Black triangle \\
	\midrule
	\cite{vander14} & 3D-HST, CANDELS& $0.3 <M_{\star} <1$ & & U-V, V-J & 0 < z < 2 & cyan dashed curve  \\
	model & & $0.3 < M_{\star} > 1$ &  & $sSFR<1$ & 0<z<2  & black dashed lines\\
	\bottomrule
\end{tabular}
\label{tab:select}
\end{table*}

The median size of model ETGs at $z=0$ is $\sim1$~kpc lower than the SDSS data quoted by \citet{cassata11}(yellow line).
At $z\sim 1$, ETGs in model have very similar size to their counterparts in observations, either  selected by morphology \citep{vander05, vander08}(green squares) or spectral energy distribution \citep{diSergo05}(red triangles). For ETGs selected by both morphology and specific star formation rate \citep{cassata11}, the model  predictions are consistent with the observed results at the $1\sigma$ level up to redshift $z \sim 1.5$. 

In the study of \citet{vander14}(cyan dashed curve), ETGs were selected by
by color -- we used sSFR as a proxy for this selection as colous in the model are subject to additional uncertainties. The discrepency between the model and these observations decreases at higher redshift. Samples drawn from the model with the mass and morphology criteria of \citet{cenarro09}(yellow line) lie on the line defined by our fiducial selection, because the morphology cut is the same and the mass cut makes no practical difference. However, the data of \citet{cenarro09} define a significantly steeper relation, with a larger median size compared to the model at low redshift and slightly smaller median size compared to the model at $z = 2$.

Although there is some tension between the model and data at high and low redshift, the trend of size evolution of the model 
from $z=2$ to the present day is in reasonable agreement with observations. ETG samples have a smaller median size at 
$z\sim2$ by a factor $\sim1.8$. We note that, since the mass range defining ETGs is only bounded at low mass, the fact that 
more high-mass ETGs enter the sample at low redshift would cause some evolution in the `average' size of the ETG population 
even if the amplitude and slope of the size -- mass relation remained fixed. However, as the galaxy mass function truncates 
exponentially at high masses, the contribution of additional high mass galaxies to the average size measure is only of the order 
of  a few per~cent. Hence, the increase in average size must be driven by a change in the overall amplitude of the size -- mass relation with redshift.

\begin{figure}
\includegraphics[width=0.45\textwidth, clip=True]{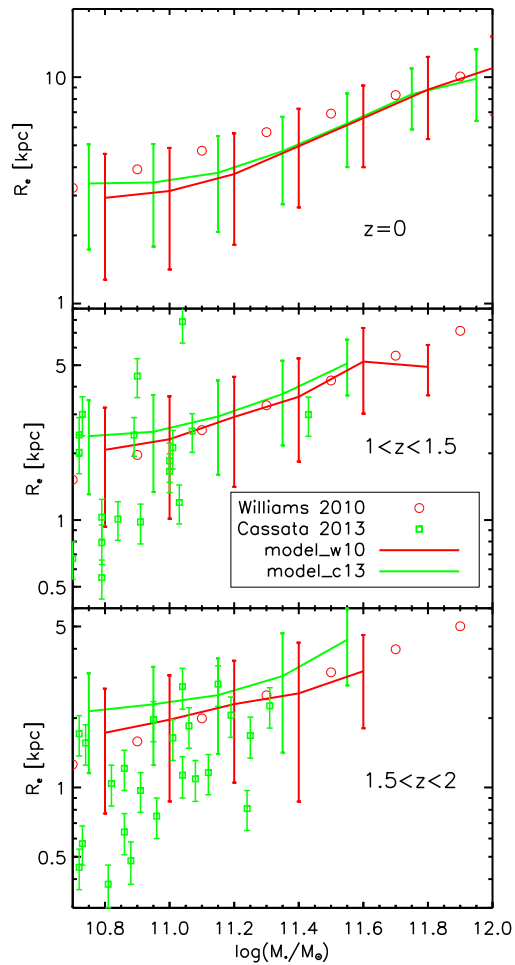}
\caption{The size -- stellar mass relation for ETGs selected at different redshifts. Red circles \protect\citep{william10} and green squares \citep{cassata13} are observational results in different redshift intervals, as indicated in each panel. Red and Green curves are the correspnond model predictions, with equivalent selections as described in the text. Error bars show the  $1\sigma$ dispersion. Model galaxies are selected at the median redshift of each redshift interval of the corresponding observations, not over the entire redshift interval.}
\label{fig:ReM}
\end{figure}

In Fig.~\ref{fig:ReM} we compare the evolution of the model size vs. stellar mass 
relation with observations in three redshift intervals. Red circles in each panel 
show measurements by \citet{william10}, who used SDSS data at $z\approx0$ and an 
updated version of the K-selected galaxy catalog of \citet{william09} for higher redshift
galaxies. ETGs are definde by sSFR$ < 0.3/t_\mathrm{H}$, where $t_{\mathrm{H}}$ is 
the age of the Universe at the corresponding redshift. We select model ETGs with the 
same criteria; the red solid curve with errors indicates their median size and its 
$1\sigma$ deviation, respectively. Note that instead of including all galaxies in 
the same redshift intervals of \cite{william10}, we select only in a narrow slice 
around the median redshift of each interval. 
For a given stellar mass, the half mass radius of ETGs selected in this way is predicted to increase by a factor of $1.6$ across this range of redshift -- this change in amplitude is substantially greater than the increase in the `typical' mass of ETGs due to the inclusion of a larger fraction of very massive galaxies at low redshift. Model predictions are broadly consistent with the \citet{william10} data at all redshifts, particularly for high mass ETGs. The agreement for low mass ETGs is somewhat better at higher redshift.

Most recently, \cite{cassata13} used the Cosmic Assembly Near-infrared Deep 
Extragalactic Legacy Survey (CANDLES) to study the size vs. stellar mass relation 
for ETGs between $1< z <3$. They selected spheroidal galaxies with 
$10^{10} < \solarmass < 10^{11.5} \solarmass$ and $sSFR < 10^{-11} \mathrm{yr^{-1}}$. 
Individual measurements for each of their galaxies are shown with green symbols in 
Fig.~\ref{fig:ReM}. For comparison we select model galaxies with $sSFR < 10^{-11}yr^{-1}$ 
and $M_{bulge}/M_{star} > 0.9$. The size -- mass relations of this selection at the 
median redshift of each interval explored by \cite{cassata13} are shown with green 
solid curves ($1\sigma$ errors). Model predictions for ETGs selected in this way are 
less obviously consistent with the observational data than in the previous case, 
particularly in the highest redshift interval where the overestimate of size at low mass is even more pronounced. We believe this is (at least partly) because the \citet{guo11} model does not take into account dissipation of energy by gas during mergers. This should make the remnants of gas-rich high-redshift major mergers considerably smaller, but will not affect the remanants of gas-poor mergers, which dominate at lower redshift.

In summary, the evolution in the `typical ETG' size in the model is driven primarily by the increasing amplitude of the size -- mass relation with redshift.

\subsection{Amplitude of the size--mass relation}
\label{subsec:bias}

In this section, we will explore the origin of the evolution in the amplitude of the ETG size -- mass relation. Two factors are relevant. Firstly, galaxies classified as ETGs at high redshift can grow in mass and size over time. The size-mass relation will evolve according to the rate of size change per unit additional mass. Secondly, galaxies may enter (or leave) the ETG population. If the median size at fixed mass for `newly formed' ETGs changes over time, the size-mass relation will evolve even if individual ETGs do not. Since the number of galaxies classified as ETGs at high redshift is only a few percent of that at $z = 0$, the latter effect could easily dominate the apparent evolution in the observed size--mass relation. 

To separate these two effects in the model, we divide ETGs into three disjoint sets, according to the redshift interval in which they are first identified as ETGs according to our fiducial selection: 

\begin{enumerate}
 \label{enu:samples}
 \item \textit{HighZ}: first identified as ETGs before $z=1.6$ 
 \item \textit{MidZ}: first identified as ETGs between $1<z<1.6$
 \item \textit{LowZ}: first identified as ETGs after $z=1$
\end{enumerate}

\begin{figure}
\includegraphics[width=0.45\textwidth]{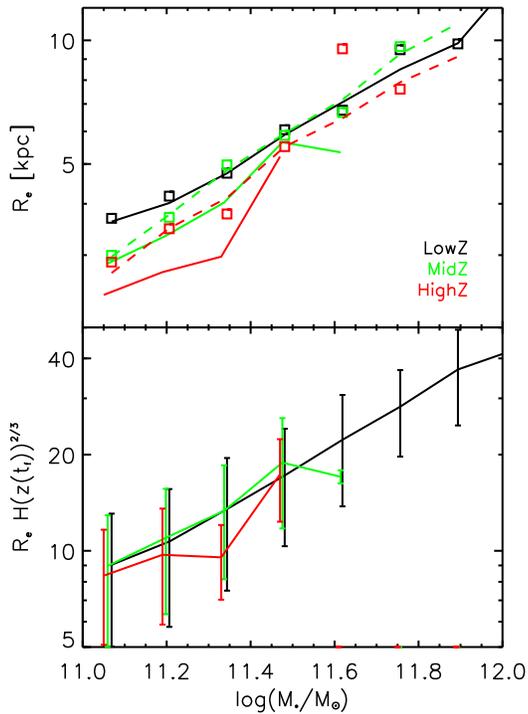}
\caption{Top panel: size -- mass relation of ETG samples
identified at different redshifts. Solid lines show the median values of our \textit{LowZ},
\textit{MidZ} and \textit{HighZ} samples. Dotted lines show the relation at $z = 0$ for descendants of the \textit{MidZ}
and \textit{HighZ} samples. Open squares show the relation defined by \textit{satellites at $z=0$} (using descendants of the \textit{MidZ}
and \textit{HighZ} samples). Bottom panel: the mean size -- mass
relation for the three samples after rescaling by their star formation time, as described in the text. Error bars show the standard error of the mean size.} 
\label{fig:ReM_evo}
\end{figure}

The top panel of Fig~\ref{fig:ReM_evo} shows the median size -- mass relation for each of these groups (red, green and black solid lines as indicated in the legend). The slope of the relation is almost independent of redshift. The amplitude increases by a factor of $\sim1.2$ between successive groups. A dashed line of the same colour shows the relation defined at by descendants of the corresponding group at $z=0$ (almost all descendants are still ETGs). Descendants of the \textit{HighZ} and \textit{MidZ} samples cover the full mass range of the \textit{LowZ} sample, showing that some galaxies have grown in mass. \textit{HighZ} galaxies remain slightly more compact than more recently formed ETGs at a fixed mass. This may also be true for \textit{MidZ} galaxies with descendants less massive than $2\times 10^{11}M_{\odot}$. 

Open square symbols in the top panel of Fig~\ref{fig:ReM_evo} also show relations for the descendants of each sample at $z = 0$, but unlike the dashed lines they only include descendants that are satellites at $z=0$. About $21\%$ of \textit{HighZ} samples at $z=0$ are satellites. The fraction is $27\%$ for \textit{MidZ} samples. It can be seen that the relations for \textit{HighZ} and \textit{MidZ} are the same regardless of whether the galaxies have become satellites or not. This is consistent with the finding of \cite{cassata11} that, among galaxies in clusters, older galaxies have smaller sizes at given stellar mass. We predict that this is true also in the field.

At a fixed mass, dark matter halos are more compact at high redshift, with the virial radius evolving as $R_{200}(z) \propto H(z)^{-2/3}$, where $H(z)$ is the Hubble parameter. In the model this redshift dependence of halo scale propagates to the sizes of newly formed galaxies, via the angular momentum conservation of infalling gas (Eq.~\ref{eqn:stardisksize}). The formation of bulges preserves the relative compactness of the system if the two progenitors are of similar age (hence size), but mergers with many more diffuse galaxies will have a `diluting' effect (Eq. ~\ref{eqn:bulgesize}). Therefore, if galaxies form most of their stars at one early epoch, their sizes should reflect the intial scale of their host halos even at $z=0$, provided that their final mass is not dominated by mergers with galaxies formed at lower redshift. 

\begin{figure}

\includegraphics[width=0.45\textwidth]{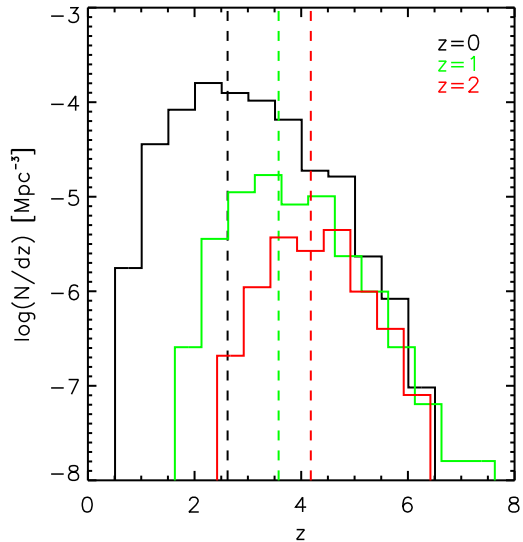}

\caption{Distribution of the star formation time of our \textit{LowZ} , \textit{MidZ} , and \textit{HighZ} ETG samples, further restricted to have $10^{11}<M_{\star}<3\times 10^{11}\solarmass$. The colours of the threee histograms correspond to those in Figure~\ref{fig:ReM_evo}. Dashed vertical lines indicate the corresponding median redshift of each sample.}

\label{fig:ft}
\end{figure}

To test this idea, we define a galaxy's formation time $t_{\rm f}$ to be the time by which half of its $z=0$ stellar mass is formed. Note that this is \textit{not} time by which half the stellar mass is assembled into a single object (often used as a defintion of galaxy `formation time' in the literature). At $t_{\rm f}$, the stars in one $z=0$ galaxy may belong to many separate galaxies. Fig.~\ref{fig:ft} shows the distribution of $t_{\rm f}$ for \textit{LowZ} , \textit{MidZ} , and \textit{HighZ} are shown with black, red, and green histograms, respectively. 
The formation time depends on stellar mass. Stars in massive ETGs assemble at earlier time than those in less massive ETGs.
To remove the mass dependence, we further restrict to those with stellar mass $10^{11}<M_{star} < 3\times10^{11}M_{\odot}$. Dashed vertical lines of the same color indicate the median formation redshift of the corresponding sample. For \textit{LowZ}, the median redshift is $2.6$, while for \textit{HighZ} the median redshift is $4.2$. The $H(z)^{-2/3}$ scaling between these two median redshifts predicts a factor of $1.4$ difference in size, close to actual the difference between the \textit{LowZ} and \textit{HighZ} samples.

To illustrate this effect, we scale the radius of each galaxy by a factor of $H(z(t_{\rm f}))^{2/3}$ and recompute the size mass relation for each sample. The results are shown in the bottom panel of Fig.~\ref{fig:ReM_evo}. Solid curves with error bars show the mean size and the standard error on the mean size. The scaled size -- mass relations overlap with each other for the \textit{LowZ} and \textit{MidZ} samples. At $\sim 10^{11}M_{\odot}$, the scaled size of \textit{HighZ} is the same as those for the \textit{LowZ} and \textit{MidZ}. This implies that most of the change in the ampltidue of the relation between these samples can be explained by the evolution of the scale of dark matter halos in which their stars form. The rescaled sizes are systematically lower for the \textit{HighZ} sample (although the statistics are poor; the relations are in marginal agreement taking into account the error on the mean values.

\section{Formation and evolution of ETGs}
\label{sec:grow}

The last section demonstrated that evolution in the characteristic star formation time of galaxies classified as ETGs largely determines the amplitude of the ETG size -- mass relation at different redshifts. In this section we explore the drivers of evolution in this relation, namely changes in the sizes and masses of galaxies over time.

\subsection{Formation of ETGs}
\label{subsec:growmass}

\begin{figure}
 \includegraphics[width=0.45\textwidth]{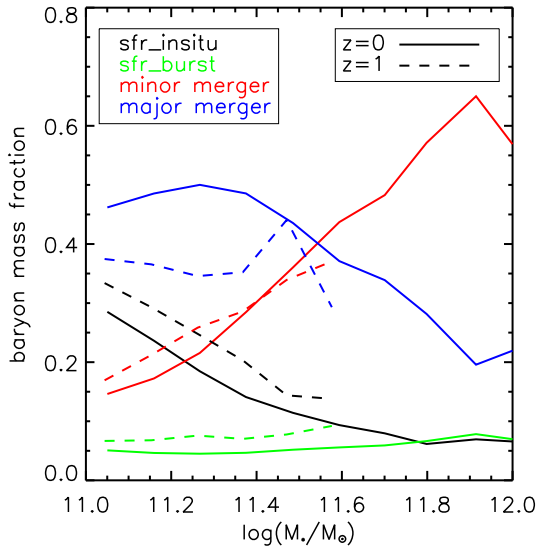}
 \caption{Stellar mass growth via different mechanisms as a function of stellar mass. Solid and dashed curves are for the \textit{LowZ} and the \textit{MidZ}, respectively. Black, green, red and blue curves are for the growth via in situ star formation, star burst, minor and major mergers, respectively.}
 \label{fig:baryonfrac}
\end{figure}

In Fig.~\ref{fig:baryonfrac} we sum the fractional contribution of in situ star formation, starbursts, minor and major mergers to mass growth along the main branch of the galaxy merger tree, for \textit{LowZ} (solid) and \textit{MidZ} (dashed) samples. This is done for every galaxy in each sample in bins of stellar mass at $z=0$; we then plot the median contribution for each mass bin.  Among `in situ' star formation processes, we separate stars formed in merger-induced starbursts (black lines) from stars formed in quiescent disks (green). We separate accrered stars according the mass ratio of their progenitor, using the minor/major merger criterion (red/blue respectively).

The contribution from mergers dominates the growth in stellar mass at all  masses. For the \textit{LowZ} sample, below $3\times10^{11} \mathrm{M_{\odot}}$, major mergers contribute a larger fraction of mass than minor mergers, while at higher masses, minor mergers dominate. The contribution from minor mergers  increases with increasing stellar mass, while contribution from major mergers deceases. The \textit{MidZ} sample does not include galaxies much beyond this transition mass; at lower masses the same trends are apparent, although major mergers become slightly less important relative to quiescent in situ star formation.  In situ formation accounts for $35\%$ and $40\%$ of the mass in the \textit{LowZ} and \textit{MidZ} samples respectively at low masses, and this contribution decreases with increasing galaxy mass. This limited contribution is not surprising in the context of the ETG selection, which, in the model, isolates objects that are dominated by massive spheroids and/or low star formation rates. The contribution from starbursts is only $5\%$ and $7\%$ for the \textit{LowZ} and \textit{MidZ} samples, respectively, over all masses shown. This agrees with previous findings that most mergers that $ETG$s have experienced were gas poor \citep{naab09,oser12,Hopkins2009}. This is reassuring, because in this regime Eq.~\ref{eqn:bulgesize} is a reasonable approximation to size growth during mergers \citep{cole2000}.

\begin{figure}
\includegraphics[width=0.45\textwidth]{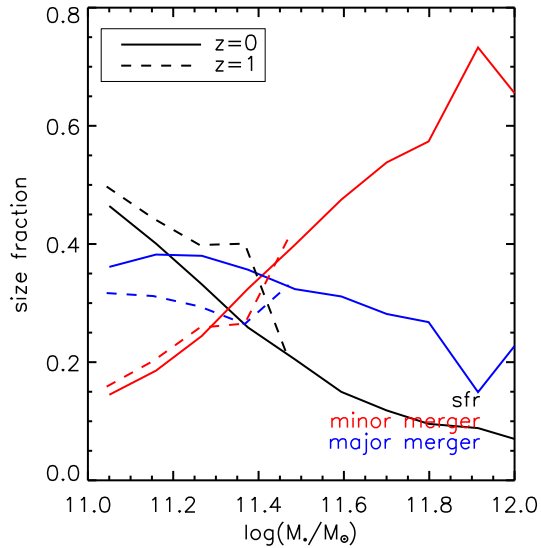}
\caption{The growth of effective radius through different mechanisms as a function of stellar
mass. All curves show median values. Black, red and green curves are 
contributions from in-situ star formation, minor merger and major mergers respectively.
Solid and dashed lines show the results of ETGs at redshift z=0 and $z\sim1$.}
\label{fig:sizefrac}
\end{figure}

In Fig.~\ref{fig:sizefrac} we compare the contribution of each of these galaxy-building processes to the change in galaxy size. In the model galaxy mass always increases, but the net change in size for individual galaxies can, in principle, be negative. The average change for galaxies in our ETG samples is positive, however. We combine the contribution of major or minor merger and the starburst it intrigerS and treat them together.
Disk instability has neglected effect on size since typically only $3\%$ of the final bulge mass is involved.

The relative contributions of major and minor mergers to size change are very similar to their contribution to the stellar mass. The fraction of size growth attributable to major mergers decreases with final galaxy mass, while the fraction attributable to minor mergers increases. Interestingly, in situ star formation plays a more important role in the size growth of the \textit{LowZ} sample compared to that in the mass growth, being responsible for to up to 50\% of the net increase in size of galaxies with final mass $\sim10^{11}M_{\odot}$. The importance of in situ star formation decreases very rapidly with increasing stellar mass, as expected from its neglibile contribtion to the stellar mass in this regime. Mergers domiante the growth of the ETG size. We draw similar conclusions from the \textit{MidZ} sample, in line with the relative differences in the contributions to the stellar mass growth between the two samples.

\begin{figure}
 \includegraphics[width=0.5\textwidth]{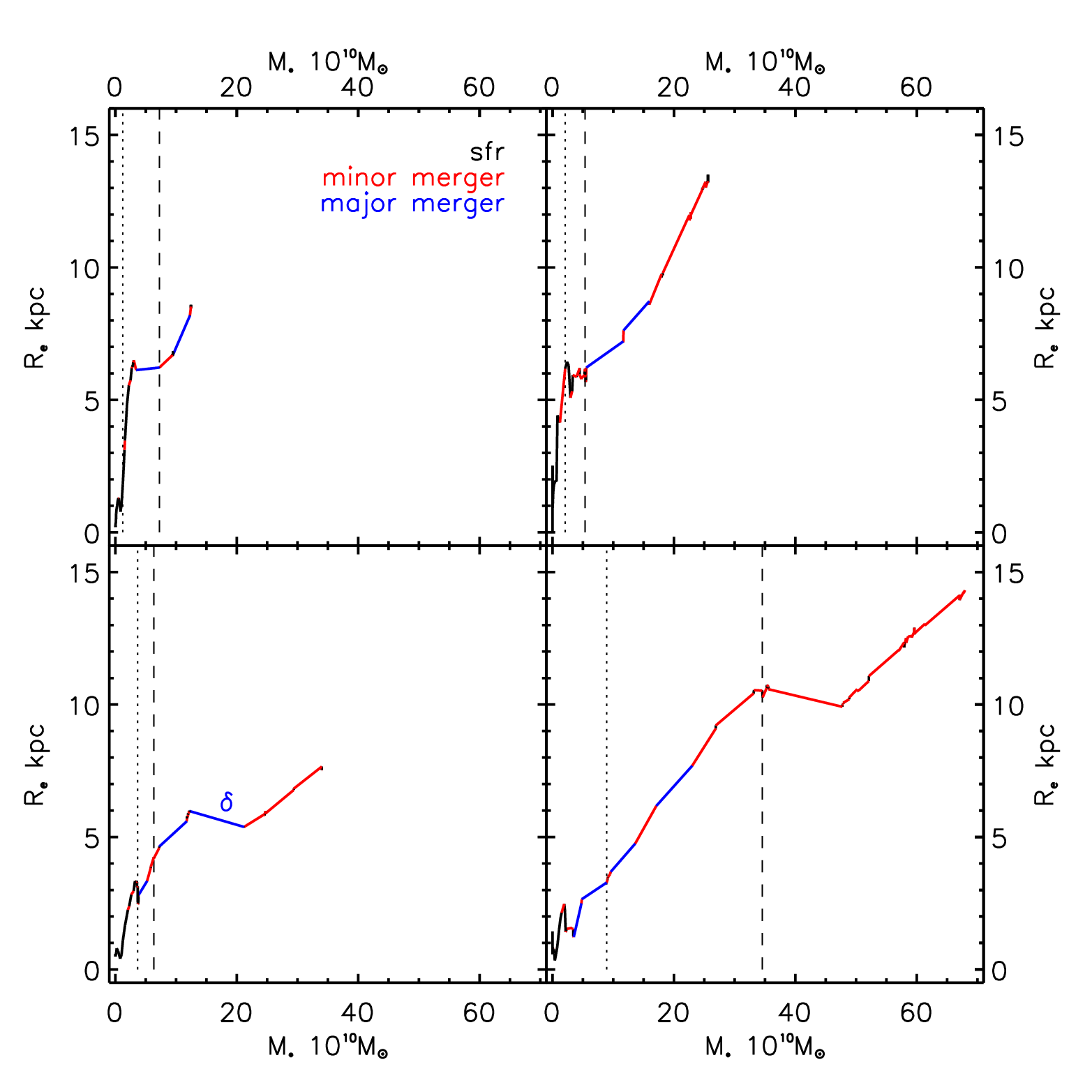}
\caption {Examples of trajectories in the mass--size diagram for individual ETGs randomly selected from our sample. Line colours indicate the mode of growth in each size change (black: star formation; blue: major mergers; red: minor mergers). Vertical dotted and dashed lines indicate the masses corresponding to $z=2$ and $z=1$ respectively.}
\label{fig:ETGexamples}
\end{figure}

In Fig.~\ref{fig:ETGexamples} we show how four individual $z=0$ ETGs in the model evolved, in order to understand better the average behaviour seen in previous figures. Each time one of these galaxies changes in size and mass, we plot a  vector joining the initial and final positions in the size -- mass plane. Mass increases monotonically, but size can increase or decrease. The colour of each vector corresponds to the mechanism responsible for the change. At low masses (high redshifts) our example trajectories are dominated by star formation (black), which contributes little to the final mass but induces rapid fluctuations in size. After $z=2$ (dotted vertical lines), however, merging dominates. Consisten with the average behaviour, major mergers (blue) dominate in our lowest mass example and minor mergers (red) become more significant at higher final masses.

There are several notable features in the trajectories shown in Fig.~\ref{fig:ETGexamples}. For example, there is no clear distinction in the typical slope 
(size change per unit additional mass) of the line segments corresponding to major and minor mergers. This is in part because the formula used to compute 
size change (Eqn.~\ref{eqn:bulgesize} ) explicitly includes the mass ratio -- it does not include a sharp boundary between the two classes of merger. In most cases the 
masses contributed by 'minor' mergers are almost as large as those contributed by major mergers, suggesting that they are not all that 'minor' and the 
distinction is somewhat artificial in this context. Another reason for the lack of a clear distinction is also related to the perhaps surprising result 
that mergers can have near-zero net size change and, in some cases, even make the remnant more compact than the primary progenitor. In the model this is 
not due to the dissipation of interaction energy by gas (which is not included) nor to the nuclear starburst (which contributes only a small amount of mass). 
Rather it is the result of mergers between a diffuse primary and a more compact secondary. Examination of Eqn.~\ref{eqn:bulgesize} shows that this is readily acheived with 
mergers of moderate mass ratio. An example shown in Fig.~\ref{fig:ETGexamples}, where the blue $\delta$ marks a 
case where size decreaseis after a major merger. Before the merger, the effective radius of the two progenitors are 
$5.9$~kpc and $1.7$~kpc. The corresponding stellar masses are $12.2\times 10^{10}\solarmass$ and $8.8\times 10^{10}\solarmass$. 
After merger, the size of the remnant is $5.1$~kpc.

\subsection{Evolution of high redshift ETGs}
\label{subsec:minor}

\begin{figure*}
 \includegraphics[width=0.8\textwidth]{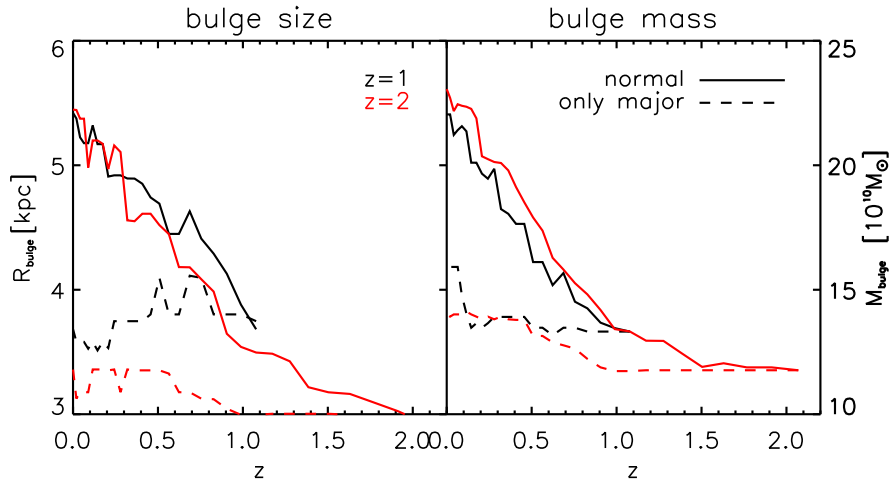}
\caption {Median bulge half mass radius (left) and median bulge mass (right) as a function of
redshift. Black and red curves correspond to our \textit{MidZ} and \textit{HighZ} samples, respectively. Solid lines show the contributiosn of all mergers (including associated starbursts) and dashed lines the contributions from major mergers alone.}
\label{fig:ETG2}
\end{figure*}

In the previous section we considered the contributions to size and mass growth since formation for all galaxies 
classified as ETGs at a particular epoch. This simulation also allows us to study the related (but not identical) 
question of how ETG galaxies identified at a particular epoch subsequently evolve. 
Many authors have considered this question. \citet{cenarro09} found the velocity 
dispersion of ETGs barely evolved since redshift $z\sim 2$ and thus concluded that minor mergers are 
be responsible for the growth in size. Using hydrodynamical simulations,
\citet{naab09} and \citet{oser12} also found that minor mergers are the main cause 
of ETG size evolution. Here we examine the causes of size growth in the \textit{HighZ} and \textit{MidZ} samples draw from the 
the semi-analytical galaxy formation model.

Fig.~\ref{fig:ETG2} shows how the average size and mass of \textit{HighZ} ETGs (red
solid lines) and \textit{MidZ} ETGs (black solid lines) evolve to lower redshift.  As
these galaxies may re-grow disks, we consider only the bulge component. The
evolution of the \textit{MidZ} sample more or less tracks that of \textit{HighZ} ETGs,  even
though most of the \textit{MidZ} galaxies were not ETGs at $z\sim 2$.  We see that the
median bulge mass of \textit{HighZ}ETGs increases by a factor of $2$ and the median
size increases by a factor of $1.9$. The fractional changes in the \textit{MidZ}
sample are similar ($1.7$ in mass and $1.5$ in size).

We then switch off minor mergers and disk instabilities after the
redshift of selection: stars accreted from satellites during minor mergers are
added to the disk rather than the bulge, as are stars formed in bursts
associated with these mergers.  The resulting size and mass growth is
considerably smaller in this case (dashed curves); a factor of $1.2$ in mass
and $1.1$ in size between $z \sim 2$ and $z = 0$.  Major mergers are therefore
only responsible for weak size evolution from $z=2$; from $z=1$ they make
almost no net contribution to mass growth and even appear to slightly decrease
the average size.  Clearly, the growth of $z\sim 2$ ETGs in the model is driven
by minor mergers, in line with previous findings from semi-analytic models
\citep{delucia2007} and hydrodynamical simulations \citep{naab09,oser12}.

\section{Conclusion}
\label{sec:conclusion}

We use an up-to-date semi-analytic galaxy formation model \citep{guo11} to
study the size evolution of massive early type galaxies (\etgs). We find that
the typical half light radius of model \etgs selected at $z = 0$ is a factor of
1.8 larger than that of galaxies selected in the same way at $z = 2$. This
finding is broadly consistent with several recent observational results. 

This increase in the typical size of the ETG population can be attributed to
two factors related to the fact that different galaxy populations are selected
by our fiducial \etg criteria at different redshifts (the abundance of \etgs
defined by these criteria increases by two orders of magnitude between $z=2$
and $z=0$, so the majority of present-day \etgs cannot have been \etgs at
higher redshift).

Firstly, since these criteria do not impose an upper limit on stellar mass,
evolution in the shape of the high-mass end of the \etg stellar mass function
results in an increasingly large fraction of extremely massive and extended
galaxies entering the \etg sample at lower redshifts. However, the median mass
increases only by $25\%$ as a result of this effect, which is far from enough
to account for the factor of $1.8$ difference in the typical \etg size.

The second and more significant factor is a `real' difference in the size --
stellar mass relation of galaxies selected as \etgs at these different
redshifts. We find that the \citet{guo11} model reproduces the slope and
amplitude of the observed relation from $z \sim 2$ to $z = 0$ at the $1\sigma$
level. This motivates us to explore the origin of this evolution in the model.
In the standard $\Lambda$CDM cosmology, dark matter halos forming at earlier
times are more concentrated. Our model for the initial sizes of gas discs based
on the conservation of angular momentum translates this into a smaller average
scale for galaxies formed at high redshift.  We can demonstrate this by scaling
the radii of ETGs according to the time when half of their stars formed; with
this rescaling, \etg samples selected at different redshifts lie on roughly the
same median size-mass relation. Hence the increase in amplitude of the relation
at lower redshift is mostly explained by increasing numbers of more recently
formed galaxies entering the \etg sample.

Mergers, stellar accretion and further in situ star formation during the low
redshift evolution of \etgs preserve this formation time dependence, to a
greater or lesser extent. Mergers always dominate over star formation in the
late-time mass and size growth of \etgs.  Minor mergers play a more important
role only for ETGs at  $M_{\star} > 3\times10^{11}M_{\odot}$. Merger-driven
starbursts contribute only about $5\%$, in line with the fact that most mergers
are gas poor after $z \sim 1$. In situ star formation contributes more to size
growth than it does to stellar mass growth, though in both cases, it is
sub-dominant compared to mergers. We find that ETGs selected at high redshifts
grow mainly via minor mergers to their present day configuration. Bulge mass
and size grows by a factor of $2.0$ and $1.9$ from $z \sim 2$ to $z =0$,
respectively. Our study of individual galaxies in the model highlights the fact
that individual mergers and star formation events can also decrease their size.

We note that the model used here does not take into account the gravitational
energy dissipated by gas processes during mergers, which could reduce the size
of the remnant. This is most important at higher redshift where gas rich
mergers are more common, reflected in the relatively larger size of the model
predictions when compared to the data at $z \sim 2$ \cite{shankar13}.

Most ETGs identified at high redshift evolve into central galaxies at the
present day (although most central galaxies today were \textit{not} \etgs at
high redshift). It is interesting to study the progenitors of ETGs identified
at different redshifts and their relation with other high redshift populations,
such as extremely red galaxies and Lyman-break galaxies. This will be done in a
companion work in the near future (Xie et al. in prep).

\section*{acknowledgments}
We acknowledge support from NSFC grants (Nos 11143005, 11133003 and 11303033 ) and  the Strategic Priority Research Program ``The Emergence of Cosmological Structure''  of the Chinese Academy of Sciences (No. XDB09000000).  QG also acknowledges a Royal Society  Newton International Fellowship. LR acknowledges the support from Youth Innovation Promotion Association of CAS.  LG also acknowledges the {\small MPG} partner Group family,  and an {\small STFC} Advanced Fellowship, as well as the hospitality of the Institute for Computational Cosmology at Durham University.

\bibliographystyle{mn2e}

\setlength{\bibhang}{2.0em}
\setlength\labelwidth{0.0em}

\bibliography{size}

\end{document}